\documentstyle[preprint,aps]{revtex}

\begin{document}

\draft

\title{ Bounds for the phonon-roton dispersion \\
in  superfluid $^4$He }

\author{J. Boronat$^a$, J. Casulleras$^a$, F. Dalfovo$^b$,
S. Moroni$^c$, and S. Stringari$^b$}

 \address{$^a$ Departament de F\'\i sica i Enginyeria Nuclear, Campus
      Nord B4-B5, \\  Universitat Polit\`ecnica de Catalunya, E-08028
      Barcelona, Spain }

\address{$^b$ INFM, Dipartimento di Fisica, Universit\`a di Trento
      I-38050 Povo,  Italy }

\address{$^c$ Istituto Nazionale di Fisica della Materia, Laboratorio FORUM,
\\ Scuola Normale Superiore, Piazza dei Cavalieri 7, I-56126 Pisa }

\date{February 28, 1995}
\maketitle

\begin{abstract}
The sum rule approach is used to derive upper bounds for the
dispersion law $\omega_0(q)$ of the elementary excitations of a
Bose superfluid. Bounds are explicitly calculated for the phonon-roton
dispersion in superfluid $^4$He, both at equilibrium ($\rho=0.02186$
\AA$^{-3}$) and close to freezing  ($\rho=0.02622$ \AA$^{-3}$).
The bound $\omega_0(q) \le 2S(q)\mid\chi(q)\mid^{-1}$,
where $S(q)$ and $\chi(q)$ are the static structure factor and density
response respectively, is calculated microscopically for
several values of the wavevector $q$.  The results provide a significant
improvement with respect to the Feynman approximation $\omega_F(q)=
q^2(2mS(q))^{-1}$. A further, stronger bound, requiring the additional
knowledge of the current correlation function is also investigated.
New results for the current correlation function are presented.
\end{abstract}


\narrowtext

\section{Introduction}
\label{sec:intro}

The microscopic investigation of the {\em dynamic} behavior
of superfluid $^4$He has been the object of extensive theoretical
work in the past starting from the pioneering works by Bijl and
Feynman \cite{Fey54} (see, for instance, Ref.~\cite{Gri93}
for an up-to-date review). Recent approaches, based on perturbation
theory with correlated functions \cite{Man86}  and on the use
of shadow variables \cite{Wu91}, have provided accurate predictions
for the dispersion law of this strongly interacting Bose system.

The purpose of this paper is to show that in  a Bose superfluid
the knowledge of relevant {\em static} properties of the system
can be used to derive useful upper bounds for the excitation spectrum,
employing a sum rule approach (for a recent discussion on sum rules
in Bose superfluids see for example Ref.~\cite{Str92}). We will show
that a key role in this context is played by the static density response
for which Diffusion Monte Carlo (DMC) calculations have recently become
available \cite{Mor92}. Another relevant quantity in this context
is the kinetic structure function, for which new DMC results will be
presented.

Explicit results for various bounds at the equilibrium density,
$\rho=0.02186$ \AA$^{-3}$, as well  as close to freezing,
$\rho=0.02622$ \AA$^{-3}$, will be given  in the first part of the
work. In the second part we will discuss in detail the behaviour
of the static response function, extending  the analysis of
Ref.~\cite{Dal92} to lower $q$'s and high pressure.

\section{Bounds for the phonon-roton dispersion}
\label{sec:part1}

The most famous estimate of the dispersion law in a Bose superfluid
was proposed many years ago by Bijl and Feynman \cite{Fey54}. The
resulting  dispersion can be written in the form
\begin{equation}
\omega_F(q) = {m_1(q) \over m_0(q)} = {q^2 \over 2m S(q)} \; ,
\label{eq:omegaf}
\end{equation}
where
\begin{equation}
m_k(q) =  \int_0^{\infty}\! d\omega \ \omega^k S(q,\omega)
\label{eq:mk}
\end{equation}
are the $k$th-moments of the dynamic structure function $S(q,\omega)$
($\hbar=1$ in this work). In deriving Eq.~(\ref{eq:omegaf}) one has
evaluated  the moment $m_1$ through the well known f-sum rule \cite{Pin66}
\begin{equation}
m_1(q)\  =\  {1\over 2} \langle [\rho_{-{\bf q}},[H,\rho_{\bf q}]] \rangle
\ =  \ N  {q^2 \over 2m}
\label{eq:m1}
\end{equation}
holding for systems of particles interacting with velocity independent
potentials.  The brackets $\langle \dots \rangle$ denote ground
state averages, while $H$ is the $N$--body Hamiltonian of the
system
\begin{equation}
  H =  - \sum_{i=1}^{N} { {\mbox{\boldmath $\nabla$}}_i^2 \over 2m} +
\sum_{i<j}^{N} V(r_{ij}) \ ,
\label{hamilto}
\end{equation}
and $\rho_{\bf q}$ is the density fluctuation operator
\begin{equation}
      \rho_{\bf q} = \sum_{i=1}^{N} e^{-i \, {\bf q} \cdot {\bf r}_i} \
\; .
\label{densoperator}
\end{equation}
The moment $m_0$ has been instead expressed in terms of
the static structure factor $S(q)$ through the equation
\begin{equation}
m_0(q) = \langle \rho_{-{\bf q}}\rho_{\bf q} \rangle  = NS(q) \; .
\label{eq:m0}
\end{equation}
Both results (\ref{eq:m1}) and (\ref{eq:m0}) have been obtained using
the completeness relation.
The static structure factor $S(q)$ is known with great accuracy both
from Monte Carlo  calculations and experimental data. In the present
work we use the Diffusion Monte Carlo results shown in Fig.~\ref{fig:sq}.

The Feynman energy (\ref{eq:omegaf}), being based on the ratio of the
two moments $m_1$ and  $m_0$, provides, at zero temperature, a rigorous
upper bound to the energy $\omega_0(q)$ of the lowest state  excited by
the density operator $\rho_{\bf k}$,
\begin{equation}
\omega_0(q) \le \omega_F(q) \; .
\label{eq:boundf}
\end{equation}
In the following we will identify the energy $\omega_0(q)$ with the one
of the elementary excitations of the system, i.e., the phonon-roton spectrum
(in a Bose superfluid this identification is exact apart from decay processes
of the elementary mode into two or more excitations \cite{Pit92}).

The bound (\ref{eq:omegaf}) reproduces exactly the phonon dispersion at
small $q$:
\begin{equation}
\omega_0(q) = cq
\label{eq:cq}
\end{equation}
where $c$ is the sound velocity. This follows from the low $q$ behavior
of the static structure factor
\begin{equation}
S(q)_{q\to 0} = {q\over 2mc}
\label{eq:sqlowq}
\end{equation}
and is the consequence of the fact that in the macroscopic regime both
the moments $m_0$ and $m_1$ are exhausted by the phonon mode.

At higher wave vectors  the Feynman bound instead overestimates
significantly the experimental  dispersion law
(see Fig.~\ref{fig:spectrum}). This is due to the occurrence of
multipair excitations, whose sthength distribution, at energy higher than
$\omega_0(q)$, turns out to be particularly important in the
determination of the energy weighted
moment. For example the experimental data of Ref.~\cite{Cow71}
at s.v.p. indicate that the roton exhausts only 1/3  of the
energy weighted sum rule, the remaining part being
associated with high energy multipair excitations.

The idea to go beyond the Feynman approximation employing a sum rule
approach was first developed many years ago by Feenberg \cite{Fee69}
with the help of the moments $m_2$ and $m_3$. Here we show that better
bounds can  be calculated using the inverse energy weighted moment
\begin{equation}
m_{-1}(q) = \int \! d\omega \  {S(q,\omega) \over \omega}
\label{eq:m-1}
\end{equation}
This moment is an ideal quantity in order to investigate the collective
properties of a Bose superfluid. In fact the factor $1/ \omega$
quenches significantly the high frequency tail of $S(q,\omega)$ where
multipair excitations are important. Furthermore the absence of single
particle excitations in the low energy part of the spectrum, typical feature
of a Bose system, makes the integral (\ref{eq:m-1})  particularly
sensitive to the contribution of the collective mode.
 For the same reason the experimental determination of $m_{-1}$, through
a direct integration of the dynamic structure factor measured by
neutron scattering, turns out to be more accurate than the one of any other
moment \cite{Cow71}.

The inverse energy weighted sum rule (\ref{eq:m-1}) is directly related to
the static density response of the system through the equation
\begin{equation}
\chi(q) = -2m_{-1}(q)
\label{eq:chi}
\end{equation}
The static response fixes the linear changes in the density induced by an
external static field interacting with the system with a potential
 of the form $H_{ext} = \lambda \rho_{\bf q}$, coupled to the density
fluctuation operator $\rho_{\bf q}$. At small $q$ it yields the
compressibility of the system
\begin{equation}
\chi(0) = - {N\over mc^2}
\label{eq:chi0}
\end{equation}
while at larger $q$ is characterized by the occurrence of a pronounced peak
(see discussion in Sect.~\ref{sec:part2}).
The static response $\chi(q)$ has been recently
calculated in superfluid $^4$He using Diffusion Monte Carlo techniques
\cite{Mor92}. These calculations reproduce the experimental data of
$\chi$ at s.v.p. with good accuracy.

The knowledge of the static response, together with the one of the
static structure factor can be used to calculate a new upper bound
for the dispersion law using the ratio
\begin{equation}
\omega_{\scriptstyle 0-1}(q) =
{m_0(q) \over m_{-1}(q)} = {2S(q) \over \mid\chi(q)\mid}
\label{eq:omega01}
\end{equation}
between the non energy weighted and the inverse energy weighted sum rules.
Since at zero temperature $S(q,\omega)=0$ for $\omega <0$, the following
inequality rigorously holds:
\begin{equation}
\omega_0(q) \le \omega_{\scriptstyle 0-1}(q) \le \omega_F(q)
\label{eq:bound01}
\end{equation}
At small $q$ also the bound (\ref{eq:omega01}) approaches the phonon
dispersion law as one can immediately  see using results (\ref{eq:sqlowq})
and (\ref{eq:chi0}). In Fig.~\ref{fig:spectrum} the new bound
is reported for several values of $q$. The improvement with respect to the
Feynman bound is significant both in the roton and in the maxon region.
 As already anticipated this improvement is the consequence of
the fact that the moments $m_0$ and $m_{-1}$, entering
Eq.~(\ref{eq:omega01}), are much  less affected by multipair
excitations with respect to the moment $m_1$. This is also true at
high pressure, as shown in Fig.~\ref{fig:spectrum_hp}.

A further improvement of the bound (\ref{eq:omega01}) can be obtained
with the help of the energy weighted  moments $m_1$ and $m_2$. In fact
one can derive the following inequality for the excitation energy
$\omega_0(q)$:
\begin{equation}
\omega_0(q) \le {1\over 2} \left[ \omega_{0-1} - \tilde \epsilon -
\sqrt{ (\omega_{0-1} -\tilde \epsilon)^2 + 4 \omega_{0-1}
\tilde \Delta } \right]
\label{eq:newbound}
\end{equation}
This bound is stronger than the bound (\ref{eq:bound01}). It
involves the knowledge of the variance
\begin{equation}
\tilde \Delta (q) = {m_1(q)\over m_0(q)} - {m_0(q) \over m_{-1}(q) }
\label{eq:delta}
\end{equation}
and of the energy
\begin{equation}
\tilde \epsilon (q) = \tilde \Delta^{-1}(q) \left[ {m_2(q) \over m_0(q)} +
\left( {m_0(q) \over m_{-1}(q) } \right)^2 - 2 {m_1(q) \over m_{-1}(q)}
\right] \ \ \ .
\label{eq:epsilonb}
\end{equation}
The latter depends not only on the moments $m_{-1}$, $m_0$ and $m_1$,
already discussed above, but also on the moment $m_2$.
The quantity $\tilde \Delta$ vanishes when $q\to 0$ since the two
energies  $\omega_{\scriptstyle 0-1}(q)$ and $\omega_F(q)$
coincide in this limit as already pointed out before. Viceversa,
the energy $\tilde \epsilon$, which represents  an average energy of
multipair  excitations, is expected to depend less critically
on $q$. Inequality (\ref{eq:newbound}) can be derived by using the
fact that  $S(q,\omega)$ vanishes for $\omega$ less than $\omega_0(q)$ in
bulk liquid $^4$He; thus the following inequality holds for any
positive $\gamma$:
\begin{equation}
\omega_0  \le { \int \! d\omega S(q,\omega) (1+\gamma \omega)^2
\over \int\! d\omega S(q,\omega) \omega^{-1} (1+\gamma \omega)^2 } \; .
\label{eq:gamma}
\end{equation}
The same inequality can be written in terms of the moments $m_k$, and
the value of $\gamma$ can be chosen in such a way to minimize the right
hand side. After some algebra one obtains inequality (\ref{eq:newbound}).
A similar procedure can be used  to derive upper and lower bounds to
the static response function (see Ref.~\cite{Dal92}), and will be
employed in Sect.~\ref{sec:part2}.

The moment $m_2$ was first explored by Feenberg \cite{Fee69}
and turns out to be proportional to the current  correlation function.
In fact one has
\begin{equation}
m_2 (q)  =  q^2 \langle J_{zq}^\dagger J_{zq} \rangle  \ \ \ ,
\label{eq:m2}
\end{equation}
where $J_{zq}$ is the $z$-component of the current density operator, and
${\bf q}$ is taken in the $z$-direction.
Using the definition of the kinetic structure function \cite{Fee69}
\begin{equation}
D(q) = {(N-1) \over q^2 \langle \psi_0 | \psi_0 \rangle }
\int \! d{\bf r}_1 \dots d{\bf r}_N
[\cos ({\bf q} \cdot {\bf r}_{12}) -1] ({\bf q}\cdot {\bf \nabla}_1
\psi_0) ({\bf q}\cdot {\bf \nabla}_2 \psi_0)
\label{eq:dq}
\end{equation}
the following expression for the moment $m_2(q)$ holds:
\begin{equation}
{ m_2(q) \over m_1(q) } = { q^2 \over 2 m} (2 - S(q))
+ {2  \over m} D(q)  \;
\label{eq:m2dq}
\end{equation}

The kinetic structure function and, hence, the moment $m_2$, has
been directly calculated using a Diffusion Monte
Carlo alghoritm. The results are shown in Figs.~\ref{fig:dq} and
\ref{fig:m2}. The structure of $D(q)$ is almost the same at the two
densities here considered;  the curve under pressure is shifted
upwards. This is just what one expects by looking at
the large $q$ behaviour of definition  (\ref{eq:dq}), which yields
$D(q) \to (2m/3)   \langle
E_K \rangle$, for $q \gg 2 \pi \rho^{1/3}$. Our values of the
mean kinetic energy $\langle E_K \rangle$ are $14.32(5)$ K and $19.57(5)$ K
at $\rho=0.02186$ \AA$^{-3}$ and $\rho=0.02622$ \AA$^{-3}$
respectively. They give an asymptotic shift of $0.3$ \AA$^{-2}$,
in agreement with the data plotted in Fig.~\ref{fig:dq}.
The curve for $D(q)$ at equilibrium density is also similar to
the one used in Ref.~\cite{Dal92}; in that case, the quantity  $D(q)$ was
obtained by Fourier transforming the results of Path Integral Monte Carlo
calculations \cite{Pol87} of the current correlation function in
${\bf r}$-space.

The microscopic results for the moments $m_0$, $m_{-1}$ and $m_2$ allow
one to calculate the bound (\ref{eq:newbound}); the results are reported in
Figs.~\ref{fig:spectrum} and \ref{fig:spectrum_hp}. All the moments entering
this analysis have  been calculated employing the Aziz potential HFDHE2
\cite{Azi79}. One notices a systematic improvement with respect to the bound
(\ref{eq:omega01}) in the whole range of wavelength from maxons to
rotons. At  equilibrium density the bounds (\ref{eq:omegaf}),
(\ref{eq:omega01}) and (\ref{eq:newbound}) yield the roton
minimum at about $17.5 K$, $11.8 K$  and $10.8 K$, respectively,
to be compared  with the experimental value $8.6K$.  The error bars in the
figures are due to statistical errors in the  calculation of $m_{-1}$  and
$m_2$.   As concerns the pressure dependence, we note that the roton
minimum shifts slightly to higher wave vectors by increasing pressure,
in agreement  with the experimental trend \cite{Sti90}.
Also the roton gap exhibits  the correct trend, being smaller
at $\rho = 0.02622$. However, the statistical error on $m_{-1}$
prevents an accurate comparison with the experimental
shift; in fact, the experimental roton gap decreases by only $1.3$ K in
the same range of pressure.

\section{Bounds for the static response function}
\label{sec:part2}

So far we have used theoretical data for the moments $m_{-1}$, $m_0$,
$m_1$ and $m_2$ in order to evaluate rigorous upper bounds for the
phonon-roton dispersion. Here we apply the formalism of Ref.~\cite{Dal92}
to evaluate upper and lower bounds to the moment $m_{-1}(q)$, and
hence to the static response function $\chi(q)$,  using $m_0$,
$m_1$, $m_2$, $m_3$, as well as the experimental phonon-roton
dispersion. This procedure will provide a check of consistency between
the available theoretical calculations for the moments $m_k(q)$, extending
the analysis of Ref.~\cite{Dal92} to lower $q$'s and to high pressure.

One can easily derive lower bounds to $m_{-1}(q)$ starting from the
inequality
\begin{equation}
\int_0^\infty  d\omega {S(q,\omega) \over  \omega}
(1 + \alpha  \omega + \beta  \omega^2)^2 \ge 0
\label{eq:alfabeta}
\end{equation}
holding for any real $\alpha$ and $\beta$. The same inequality can be
written as a lower bound to $m_{-1}(q)$. Minimization with respect to
the parameters $\alpha$ and $\beta$ provide the bounds. In particular,
by minimizing with respect to $\alpha$ with $\beta=0$ one gets the
Feynman approximation to $m_{-1}(q)$:
\begin{equation}
m_{-1}(q)  \ge m_{-1}^{F}(q) = 2 N m q^{-2} S^2(q)  \; .
\label{eq:lowerf}
\end{equation}
Minimizing with respect to both $\alpha$ and $\beta$ one gets a stronger
lower bound:
\begin{equation}
m_{-1}(q) \ge {m_{-1}^{F}(q) \over   1- \Delta(q)/\epsilon(q) } \ \ \ ,
\label{eq:lower}
\end{equation}
where
\begin{equation}
\Delta(q) = {m_2(q) \over m_1(q)}-{m_1(q) \over m_0(q)}
\label{eq:del}
\end{equation}
and
\begin{equation}
\epsilon(q) = \Delta^{-1} \left[ {m_3(q) \over m_1(q)}+({m_1(q)
\over m_0(q)})^2- 2{m_2(q) \over m_0(q)} \right] \; .
\label{eq:epsi}
\end{equation}
One notes that the quantities $\Delta$ and $\epsilon$ have the same form
of $\tilde \Delta$ and $\tilde \epsilon$ in Eqs.~(\ref{eq:delta}) and
(\ref{eq:epsilonb}), but with the index of the $k$-moments scaled by $1$.
Inequality~(\ref{eq:lower}) requires the knowledge of the cubic energy
weighted moment $m_3(q)$, which can be calculated through the Puff sum
rule \cite{Puf65}:
\begin{equation}
m_3 (q) = N \! \left[ ({q^2 \over 2 m})^3 + {q^4 \over m^2}
\langle E_K \rangle + { \rho  \over 2 m^2}
\int \! d{\bf r} \ g(r) (1 \! - \! \cos({\bf q} \! \cdot \! {\bf r}))
({\bf q} \! \cdot \! \nabla)^2 V(r) \right]
\label{eq:puff}
\end{equation}
where  $V(r)$ and $g(r)$ are the interatomic  potential and the
radial distribution function, respectively.

In a similar way, one can derive upper bounds \cite{Dal92}. One finds
\begin{equation}
m_{-1} (q) \le {m_0(q) \over \omega_0(q) } = N S(q) \omega_0^{-1} (q) \; ,
\label{eq:upperf}
\end{equation}
as well as a stronger upper bound:
\begin{equation}
m_{-1}(q) \le {m_0(q) \over \omega_0(q)} \left[ 1- {m_0(q) \over m_1(q)}
\left( {m_1(q) \over m_0(q)} -\omega_0(q) \right)^2 \left({m_2(q) \over
m_1(q)} - \omega_0(q)
\right)^{-1} \right] \; .
\label{eq:upper}
\end{equation}

The results for the above  lower and upper bounds for $m_{-1}(q)$ are
shown in Figs.~\ref{fig:chieq} and \ref{fig:chisol} at  equilibrium
density and close to freezing, respectively. To evaluate the bounds
we have used the same $m_k$ moments as in Sect.~\ref{sec:part1} and
the experimental phonon-roton dispersion for $\omega_0(q)$.
Dashed lines
correspond to the weakest bounds (\ref{eq:lowerf}) and (\ref{eq:upperf}),
while solid lines correspond to the bounds (\ref{eq:lower}) and
(\ref{eq:upper}). The latter account for the effect of multiphonon
excitations through the inclusion of higher $k$-moments. This
explains why the allowed area for $m_{-1}$, i.e. between lower
and upper bounds, is significantly reduced passing from dashed to
solid lines. Indeed the bounds (\ref{eq:lower}) and
(\ref{eq:upper}) represent a quite stringent
test of consistency between independent calculations and measurements
of $k$-moments of the dynamic structure function $S(q,\omega)$.
The available experimental data of $m_{-1}$ at equilibrium
\cite{Cow71} are consistents with the bounds. The same is true for
the Diffusion Monte Carlo data \cite{Mor92} at equilibrium and
freezing pressure. We note that the new Monte Carlo data for
$S(q)$ and $D(q)$ provides accurate bounds  even at relatively
small $q$'s, i.e, in the maxon region $0.5$ to $1.5$ \AA$^{-1}$.

\section{Conclusion}
\label{sec:conclusion}

In the first part of this work we have discussed new upper bounds for
the excitation spectrum in superfluid $^4$He. The method makes use of
basic {\em static}  properties of the system: the static structure
factor, the static response and the current  correlation function.
These quantities are now available in microscopic  {\em ab initio}
calculations with good accuracy. In particular we have used recent
Diffusion  Monte Carlo data for the static response \cite{Mor92}
and we present new results for the kinetic structure function.
The upper bounds for the phonon-roton dispersion turn out to be
rather close to the experimental values and can be calculated
at any pressure.

In the second part we have evaluated  upper and lower bounds to the
static response function using a method proposed by two of us in
Ref.~\cite{Dal92}.  The new data for the current correlation function
allows one to extend the analysis of Ref.~\cite{Dal92} to lower
values of $q$ and to high pressure. The main result is a general
consistency between the independent evaluations of the several
$k$-moments involved in the analysis, and hence of the quantities
$S(q)$, $D(q)$, $\chi(q)$ in a wide range of $q$.

\begin{figure}
\caption{
Static structure factor $S(q)$ at equilibrium density (solid line)
and at $\rho=0.02622$ \AA$^{-3}$ (dashed line) }
\label{fig:sq}
\end{figure}

\begin{figure}
\caption{
Phonon-roton spectrum at the equilibrium density, $\rho=0.02186$ \AA$^{-3}$.
Solid line: experiments \protect \cite{Don81}; dashed line: Feynman
approximation (\protect \ref{eq:omegaf}); empty circles: upper bound
$\omega_{\scriptstyle 0-1}$ defined in Eq.~(\protect \ref{eq:omega01});
solid circles: upper bound (\protect \ref{eq:newbound}). }
\label{fig:spectrum}
\end{figure}

\begin{figure}
\caption{
Same as in Fig.~\protect \ref{fig:spectrum} but at freezing pressure
($\rho=0.02622$ \AA$^{-3}$). The solid line corresponds to a smooth
interpolation between experimental data on the phonon dispersion
\protect \cite{Sve72},  up to  $1$~\AA$^{-1}$, and recent data on the
roton minimum \protect \cite{Sti90}.  }
\label{fig:spectrum_hp}
\end{figure}

\begin{figure}
\caption{
Kinetic structure function $D(q)$ at equilibrium density (empty
circles) and close to freezing (solid circles). }
\label{fig:dq}
\end{figure}

\begin{figure}
\caption{
Ratio $m_2(q)/m_1(q)$ at equilibrium density (empty circles) and close
to freezing (solid circles) }
\label{fig:m2}
\end{figure}

\begin{figure}
\caption{
Inverse energy weighted moment $m_{-1}(q)$ at equilibrium density.
Empty circles: experiments \protect \cite{Cow71}; solid circles with error
bars: Diffusion Monte Carlo calculations \protect \cite{Mor92};
dashed lines: upper and lower bounds (\protect \ref{eq:upperf}) and
(\protect \ref{eq:lowerf}); solid lines: upper and lower bounds
(\protect \ref{eq:upper}) and (\protect \ref{eq:lower}). }
\label{fig:chieq}
\end{figure}

\begin{figure}
\caption{
Inverse energy weighted moment $m_{-1}(q)$ at density
$\rho=0.02622$ \AA$^{-3}$. Points with error
bars: Diffusion Monte Carlo calculations \protect \cite{Mor92};
dashed lines: upper and lower bounds (\protect \ref{eq:upperf}) and
(\protect \ref{eq:lowerf}); solid lines: upper and lower bounds
(\protect \ref{eq:upper}) and (\protect \ref{eq:lower}). }
\label{fig:chisol}
\end{figure}

\end{document}